# What Takes the Brain so Long: Object Recognition at the Level of Minimal Images Develops for up to Seconds of Presentation Time

**Hanna Benoni (BenoniH@colman.ac.il)**
Department of Psychology
The College of Management Academic Studies
2 Elie Wiesel Street, Rishon LeZion, Israel

**Daniel Harari (hararid@weizmann.ac.il)**
Weizmann Artificial Intelligence Center
Department of Computer Science and Applied Mathematics
The Weizmann Institute of Science
234 Herzl Street, Rehovot, Israel

**Shimon Ullman (Shimon.Ullman@weizmann.ac.il)**
Department of Computer Science and Applied Mathematics
The Weizmann Institute of Science
234 Herzl Street, Rehovot, Israel

## Abstract

Rich empirical evidence has shown that visual object recognition in the brain is fast and effortless, with relevant brain signals reported to start as early as 80 ms. Here we study the time trajectory of the recognition process at the level of minimal recognizable images (termed MIRC). These are images that can be recognized reliably, but in which a minute change of the image (reduction by either size or resolution) has a drastic effect on recognition. Subjects were assigned to one of nine exposure conditions: 200, 500, 1000, 2000 ms with or without masking, as well as unlimited time. The subjects were not limited in time to respond after presentation. The results show that in the masked conditions, recognition rates develop gradually over an extended period, e.g. average of 18% for 200 ms exposure and 45% for 500 ms, increasing significantly with longer exposure even above 2 secs. When presented for unlimited time (until response), MIRC recognition rates were equivalent to the rates of full-object images presented for 50 ms followed by masking. What takes the brain so long to recognize such images? We discuss why processes involving eye-movements, perceptual decision-making and pattern completion are unlikely explanations. Alternatively, we hypothesize that MIRC recognition requires an extended top-down process complementing the feed-forward phase.

**Keywords:** object recognition; minimal recognizable images; long presentation time; sequential processing

## Introduction

Visual object recognition is a fundamental task performed frequently by the visual system. Despite its computational complexity, there is abundant empirical evidence that visual object recognition is performed in the brain quickly and effortlessly. Earliest brain signals reflecting the category of an object in the visual input were reported to be as early as 75-80 ms, using event-related potential (ERP) (VanRullen & Thorpe, 2001). In these experiments, subjects had to decide whether or not a previously unseen image, flashed for 20 ms (without masking), belonged to a target category (Thorpe,

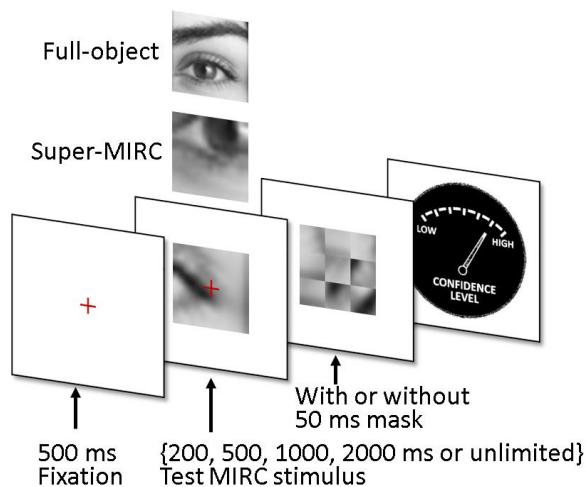

Figure 1: Behavioral experiment settings. All stimuli were greyscale images, presented on a white background. Each trial consisted of a fixation display for 500 ms, the stimulus image display, and a request to rank the confidence level in the range 1 to 7. In conditions with visual masking, a scrambled version of the test image patch was presented for 50 ms immediately following the stimulus display. In conditions with MIRC stimuli, each participant was assigned to one of nine exposure times: 200, 500, 1000, 2000 ms with or without masking, as well as unlimited time. Control conditions compared to super-MIRC patches and full-object images (shown above the main sequence).

Fize, & Marlot, 1996). In recent models of so-called 'core visual object recognition', reliable object recognition could be read out from electrophysiological recordings of primate brain activity within 70-170 ms from stimulus onset (100 ms presentation time), using large scale multi-electrode arrays placed in inferior temporal (IT) cortex (Cadieu et al., 2014; DiCarlo, Zoccolan, & Rust, 2012; Hung, Kreiman, Poggio, & DiCarlo, 2005). In another recent study, multivariate pattern classification was applied to magnetoencephalography (MEG) activity, showing peak decoding accuracy already at 102 ms for individual images, 122 ms for superordinate categorization and 170 ms for subordinate categorization (Cichy, Pantazis, & Oliva, 2014).

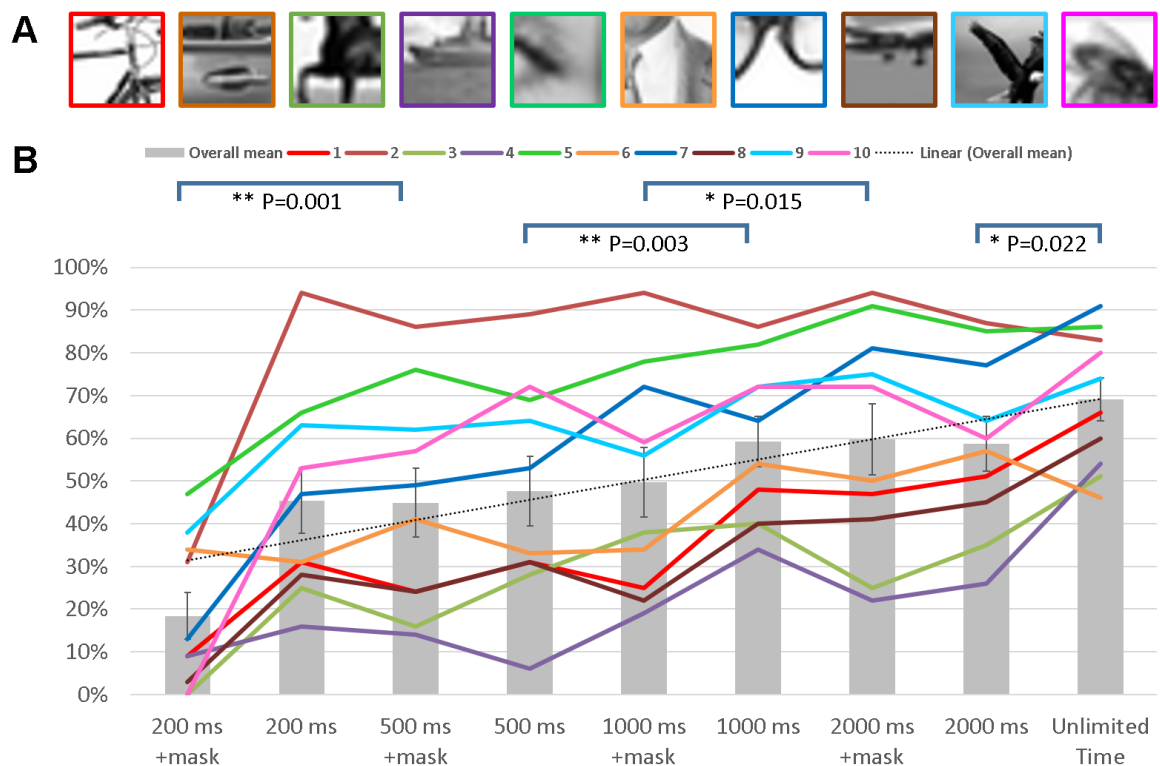

Figure 2: **(A)** Minimal recognizable images (MIRCs) from 10 object categories used as test stimuli (mean resolution of 15 image samples per MIRC). In the experiment, each image subtended 3×3 degrees of visual angle on the screen. **(B)** Recognition rates for different exposure conditions with or without masking. Colored-lines (corresponding with the frame colors around the stimuli images in (A)) indicate mean rates across subjects (see also Table 1). Bars indicate overall mean rates across subjects and MIRCs. Statistical significance of pairwise comparisons is indicated at the top of the chart. Dotted line indicates the linear trend of the overall mean.

Discovering the visual features and representations used by the brain to recognize objects, as well as the underlying computational processes, are a central challenge in the study of vision. A recent study discovered image patches of objects (at mean resolution of 15 image samples) that can be reliably recognized on their own by human observers, but are minimal in that further reduction in either size or resolution makes them unrecognizable ((Ullman, Assif, Fetaya, & Harari, 2016); examples Figure 2A). The study found that at the level of minimal recognizable images (termed MIRCs) a minute change of the image can have a drastic effect on recognition, thus identifying features that are critical for the task. It was shown that an object image is usually covered by multiple such minimal recognizable images, and that this coverage provides robustness to occlusion and distortions at the object level, because each MIRC is recognizable on its own. Interestingly, a recent fMRI study showed that MIRC images elicited enhanced brain activation in corresponding category-selective regions (fusiform face area (FFA) and occipital face area (OFA) for faces, lateral occipital cortex (LOC) for objects and para-hippocampal place area (PPA) as well as transverse occipital sulcus (TOS) for places), similar to full-object images (Holzinger, Ullman, Harari, Behrmann, & Avidan, 2019).

In this paper, we study the time trajectory of the recognition process at the level of minimal images, by controlling the display exposure (i.e. presentation) time and image masking. The results below show that recognition rate (ratio of correct responses out of the total number of responses in a trial) in the masked conditions are as low as 18% for 200 ms exposure, but gradually develops for longer exposures even above 2 seconds. MIRCs presented until response (i.e. for unlimited time) had the same recognition rates as full-object images presented for 50 ms with masking.

These findings indicate that the recognition process at the level of minimal images takes hundreds to thousands of milliseconds, which is surprisingly long compared with standard recognition of full object images. What takes the brain so long? There are several visual processes that are known to take a long time. We will discuss why these processes are unlikely to provide a full explanation of our findings, and hypothesize an explanation in terms of sequential top-down processing, which complements the feed-forward phase.

## Methods

### Participants
The participants were 445 (270 females, 175 males) undergraduate students (average age 25.2) from the College of management academic studies (Colman). All participants had normal or corrected-to-normal vision.

### Apparatus
Stimulus displays were presented on a 22-inch computer screen with a resolution of 1920×1080 pixels. Each participant was tested individually in a dimly lit room. Responses were recorded by an experimenter via a computer keyboard. A chinrest was used to stabilize viewing distance at 57.5cm from the monitor, so that 1cm on the display represented 1 degree of visual angle.

### Stimuli and Procedure
A set of 10 MIRC images from 10 different classes were used (bike, car, horse, ship, eye, suit, glasses, airplane, eagle and fly; see
Figure 2A). Control conditions compared the MIRC conditions with two related sets of stimuli (Figure 1). The first set included 10 super-MIRC images (predecessor patches in the MIRC search tree, with a few additional image samples; see (Ullman et al., 2016)), one from each of the 10 categories. The second set included the 10 full-object images (50×50 image samples), from which all of the image patches (MIRC and super-MIRC) were originated. All stimuli were greyscale images, and presented on a white background. Each trial contained a fixation display, an image display, and a question of confidence display (see experimental settings in Figure 1). The fixation display consisted of a small central red cross (RGB values of (255, 0, 0)), on a white background, subtending 0.3×0.3 degrees of visual angle. Each patch image subtended 3×3 degrees of visual angle from edge to edge. In conditions with visual masking, the masking stimulus was a scrambled version of the tested image. In the confidence level ranking display, participants were presented with the question "How confident are you?" and a visual scale of seven levels of confidence.

A group of 332 participants was assigned to the MIRCs conditions, and another group of other 113 participants was assigned to the control conditions. In the MIRC conditions, each participant was assigned to one of nine exposure time conditions: 200, 500, 1000, 2000 ms with or without masking, and unlimited time. In the control conditions, full-object images were presented for 50 or 200 ms, followed by masking. Super-MIRCs were presented for 200 ms followed by masking or for unlimited time (i.e. until response). All of the conditions consisted of 10 trials. In each trial, one image stimulus from the 10 object categories was presented. Each trial began with a 500 ms fixation display followed by the image display. The fixation mark remained on the screen throughout the image presentation. In conditions with masking, a 50 ms visual mask immediately followed the test image display. Each trial ended with a request to rank the participant's level of confidence in the trial on a scale 1 to 7. The inter-trial interval was 1500 ms.

Participants were instructed to focus on the fixation mark and were given the following instructions: "You are participating in an experiment in the area of object recognition. In this experiment, you will be presented by 10 images of objects parts, which will appear on the screen sequentially. For each image, say the name of the object you recognized. Alternately, you may say that you don't have an idea about the object identification". The time to consider the image, after presentation, was not limited. The participants responded orally. Answers were recorded by an experimenter via the 'L', 'S' and 'A' keys of the keyboard, for correct, incorrect and "Don't know" answers, respectively. No feedback was given to the participants regarding their performance. Some answers required decisions regarding the use of related terms, e.g., whether "bee" instead of "fly" would be accepted. The decision was based on the WordNet hierarchy (Miller, 1995). We allowed sibling terms that have the same direct parent (hypernym) or a common ancestor two levels up. For instance, "bee" was accepted as a label for "fly", but "spider" was not. Object-part names were accepted if they correctly labeled the partially visible object in the test image (e.g., "wheel" for bicycle or "tie" for suit).

The experiment and procedures were approved by the institutional review boards of the Weizmann Institute of Science, Rehovot, Israel, and by the Ethics Committee in the College of management academic studies (Colman), Rishon-Lezion, Israel. All participants gave informed consent before starting the experiments.

## Results

### Recognition of MIRC Stimuli
**Limited Presentation Time.** A 2X2 repeated measures ANOVA was conducted on recognition rates with masking (present vs. absent) and presentation time (200, 500, 1000 and 2000 ms). Mean recognition rates for each experimental condition are shown in Figure 2B and Table 1. The results show that the main effect of masking is significant ($F(1,9)=36.19$, $p<0.001$), hence recognition of masked presentations (M=43%) was lower compared to unmasked

Table 1: MIRC recognition rates (in percent) at increasing presentation times

| Condition | 200ms + mask | 200ms | 500ms + mask | 500ms | 1000ms + mask | 1000ms | 2000ms + mask | 2000ms | Unlimited time* |
|---|---|---|---|---|---|---|---|---|---|
| Number of subjects | N=32 | N=32 | N=37 | N=36 | N=32 | N=49 | N=32 | N=47 | N=35 |
| MIRC_1 (bike) | 9 | 31 | 24 | 31 | 25 | 48 | 47 | 51 | 66 |
| MIRC_2 (car) | 31 | 94 | 86 | 89 | 94 | 86 | 94 | 87 | 83 |
| MIRC_3 (horse) | 0 | 25 | 16 | 28 | 38 | 40 | 25 | 35 | 51 |
| MIRC_4 (ship) | 9 | 16 | 14 | 6 | 19 | 34 | 22 | 26 | 54 |
| MIRC_5 (eye) | 47 | 66 | 76 | 69 | 78 | 82 | 91 | 85 | 86 |
| MIRC_6 (suit) | 34 | 31 | 41 | 33 | 34 | 54 | 50 | 57 | 46 |
| MIRC_7 (glasses) | 13 | 47 | 49 | 53 | 72 | 64 | 81 | 77 | 91 |
| MIRC_8 (airplane) | 3 | 28 | 24 | 31 | 22 | 40 | 41 | 45 | 60 |
| MIRC_9 (eagle) | 38 | 63 | 62 | 64 | 56 | 72 | 75 | 64 | 74 |
| MIRC_10 (fly) | 0 | 53 | 57 | 72 | 59 | 72 | 72 | 60 | 80 |
| **Overall** | **18** | **45** | **45** | **48** | **50** | **59** | **60** | **59** | **69** |

* Until response

presentations (M=53%). The main effect of presentation time is also significant ($F(3,27)=32.44$, $p<0.001$). Further paired comparisons show that increasing presentation times, increase recognition rates. Recognition rates for exposure of 200 ms (M=32%) are significantly smaller than for exposure of 500 ms (M=46%; $F(1,9)=15.23$, $p=0.004$). The rates for exposure of 500 ms are smaller than for 1000 ms (M=54%; $F(1,9)=15.66$, $p=0.003$), and the recognition rates of stimuli presented for 1000 ms are marginally smaller than when presented for 2000 ms (M=59%; $F(1,9)=4.194$, $p=0.071$). Interestingly, the interaction between masking and presentation time is also significant ($F(3,27)=8.70$, $p<0.001$). Further analyses revealed the nature of this interaction. The differences of recognition rates between masked and unmasked conditions are much higher in 200 ms presentations compared to the longer presentation times ($F(1,9)=10.92$, $p=0.009$). Furthermore, the differences of recognition rates between masked and unmasked conditions are higher in 1000 ms presentations compared to the 2000 ms presentation times ($F(1,9)=8.61$, $p=0.017$).

Since it was shown that the sensory system can sustain the visual input well beyond brief presentation time in unmasked conditions (e.g. 1 second in (Sperling, 1960)), we further investigated the increasing recognition rate with the increase in presentation time specifically in the masked conditions. Pairwise comparisons in the masked conditions (Fig. 2B) show significant increase between 200 ms (M=18%) to 500 ms (M=45%; $t(9)=4.559$, $p=0.001$) and between 1000 ms (M=50%) to 2000 ms (M=60%; $t(9)=3.012$, $p=0.015$). Recognition is practically lost for exposure times less than 200 ms (M=0 for 50 ms; M=3% for 100 ms). A one-way repeated measures ANOVA analysis showed a highly significant effect of exposure time on the recognition rate ($F(4,36)=27.130$, $p<0.001$). This effect was found to be as significant also in a similar analysis of the unmasked conditions ($F(1,9)=22.125$, $p=0.001$). Interestingly, pairwise comparisons in the unmasked conditions show significant increase between 500 ms (M=48%) to 1000 ms (M=59; $t(9)=3.974$, $p=0.003$) and between 2000 ms (M=59%) and unlimited time (M=69%; $t(9)=2.773$, $p=0.022$).

**Unlimited vs. 2000 ms Presentation Time.** Although recognition is already obtained for 1000 ms exposure time (M> 50%), the recognition rate goes higher for longer presentation times, and was found to be significantly higher for unlimited exposure time (response terminated displays) compared to 2000 ms ($t(9)= 2.74$, $p=0.023$).

**Correlation between Recognition Rates in Short and Long Presentation Times.** To examine whether recognition of MIRCs in short presentation times (200 ms with masking) predicts recognition in long presentation times (unlimited), we applied the Spearman rank correlation coefficient test. The analysis revealed that the correlation between this two conditions is not significant ($r_s=0.748$; $p=0.353$). However, it should be noted that in presentations of 200 ms with masking, none of the images was recognized. Therefore, we conducted an additional analysis with presentations of 200 ms without masking (four images were recognized in this condition) and unlimited time. The analysis showed that the correlation between these two conditions is significant ($r_s$=0.329; p=0.013). Thus, images that are recognized better in short presentation times are also recognized better in long presentation times.

**Laboratory vs. AMT Results in Unlimited Presentation Time.** The image patches of MIRCs and Super-MIRCs that have been used at the laboratory experiment have been tested earlier using the Amazon Mechanical Turk (AMT) online platform with unlimited presentation time. We compared the recognition rates between the AMT and the Laboratory experiments, to validate the AMT platform as a tool.

Although the mean recognition rate of MIRCs at the lab in unlimited time (M=69%) was lower than the mean recognition rate on AMT (M=80%), this difference is only marginally significant (t(9)=2.024, p=0.074). A comparison of recognition rates of super-MIRCs shows that the difference between mean recognition rate in the lab (M=93%) and AMT (M=95%) is insignificant (t(9)=1.784, p=0.108). These comparisons suggest that AMT may be a valid and reliable tool to test recognition rates of images. Nevertheless, the tendency of higher recognition rates of images on AMT may reflect a characteristic bias that need to be considered by researchers when using this platform. This tendency was found also in previous studies (Buhrmester, Kwang, & Gosling, 2011; Hauser & Schwarz, 2016), but may be also accounted in our experiment to the fact that stimuli viewing size in the lab was fixed and possibly not optimal.

## Recognition of Full-Object and Super-MIRC Stimuli

**Recognition of Full-Object Image Stimuli** at 50 ms exposure time followed by masking (M=73%) was found to be equivalent to the recognition of MIRC stimuli in unlimited exposure time (M=69%; t(9)=0.825, p>0.1). Nevertheless, the level of confidence in correct responses was significantly higher in the condition of full objects presented for 50 ms with masking (M=5.7) compared to the condition of MIRCs presented for unlimited time (M=4.9; t(9)=2.55, p=0.031). This may suggest that restricting information in time (brief presentation of full objects) may be essentially different in some aspects from restricting information in space (MIRCs), even if it is considered similarly difficult.

**Comparison between Full-Object and Super-MIRCs.** Full-object images that are completely recognizable in unlimited exposure time (M=100%), were also found to be fully recognizable for 200 ms exposure time followed by masking (M=99%). Interestingly, super-MIRCs, yield very high recognition rates at unlimited exposure displays (M=93%), but are hardly recognizable when presented for 200 ms followed by masking (M=34%).

## Additional Results
**Level of Confidence.** The average level of confidence when recognizing MIRCs (M=4.97) was significantly higher compared to the confidence level when offering alternative (false) interpretations (M=2.82; t(313)= 30.218, p<0.001).

**Gender.** Men and Women recognized MIRCs equally (M=50%, M=51% respectively, t(327)=0.186; p=0.853). Nevertheless, men were significantly more confident in their correct responses (M=5.1), compared to women (M=4.8; t(324)=2.627; p=0.009).

## Discussion: What Takes the Brain so Long?

Behavioral evidence shows that single object recognition in natural images can usually be accomplished following a short presentation time, e.g. 100 ms in DiCarlo et al. (2012), or 50 ms for full-object images in the current study. Physiological evidence using EEG (VanRullen & Thorpe, 2001), or readout from IT neurons (DiCarlo et al., 2012; Kar, Kubilius, Schmidt, Issa, & DiCarlo, 2019) is consistent with this view. In contrast, the current study shows that recognition of minimal images (Ullman et al., 2016) takes a surpassingly long time to reach the same recognition levels that full object images reach in 50 ms presentation time.

A number of possible processes have been implicated in past studies with the relatively long processing of visual information. One is the use of eye movements in search for relevant information (Itti & Koch, 2000; Tsotsos et al., 1995). Since high-resolution vision is available only in a small region of central vision, successive fixations may be required to extract relevant visual features from multiple locations. This requirement is unlikely to play a major role in our case since the MIRC images were small (down to 1deg) and presented in central vision.

Another possible process is the integration of perceptual evidence over time to reach a decision. For instance, detecting the direction of dots moving coherently (W T Newsome & Paré, 1988) increases in time as the fraction of coherently moving dots decreases (Kiani & Shadlen, 2009; William T. Newsome, Britten, & Movshon, 1989; Shadlen & Newsome, 2002). The required presentation time can be modelled for binary decisions by a 'drift diffusion to boundary' process that integrates information over time (Palmer, Huk, & Shadlen, 2005; Ratcliff, 1978). Other studies applied the model also to static visual stimuli (Ratcliff, Hasegawa, Hasegawa, Smith, & Segraves, 2007). However, this is not directly applicable to our case, where the stimulus is unchanged and available at the input, and the decision is to be made from thousands of possible alternatives.

An additional process shown to prolong recognition time is pattern completion across occluded image regions (Tang et al., 2014, 2018). The stimuli used in the current study were un-occluded, and the presentation times required for recognition were significantly longer than in (Tang et al., 2018).

Based on (Ben-Yosef, Assif, & Ullman, 2018), we suggest that the likely process requiring long presentation time is associated with the identification of internal semantic parts within the MIRC image. Although minimal images are 'atomic' in the sense that any reduction makes them unrecognizable, human observers can recover rich internal structures within each minimal image (e.g. the tie, shirt, jacket in the torso MIRC, in Fig 2A). The recovery of internal parts is modelled in (Ben-Yosef et al., 2018) as a prolonged iterative process combining bottom-up and top-down components. The iterative process is used to compare the internal structure of a stored object model (or models) with the image. In contrast with MIRCs, full object images contain redundant information from different regions of the same object, allowing reliable classification following a brief feed-forward pass. We hypothesize that even with full object images, perceiving small internal parts may require top-down processing and a longer presentation time. The difference between full and minimal images suggests that MIRCs can provide particularly efficient stimuli for studying brain mechanisms involved in top-down visual processing since they provide relatively weak feed-forward information, and are highly dependent on subsequent top-down processes. Computationally, MIRCs can be useful to study limitations of current deep neural networks and possibly increase their robustness to small image variations (Linsley, Eberhardt, Sharma, Gupta, & Serre, 2017; Srivastava, Ben-Yosef, & Boix, 2019).

## Acknowledgments

Supported by EU Horizon 2020 Framework 785907, ISF grant 320/16 and the Robin Chemers Neustein Artificial Intelligence Fellows Program.